# ADAPTIVE MODELING OF URBAN DYNAMICS DURING EPHEMERAL EVENT VIA MOBILE PHONE TRACES

Suhad Faisal Behadili[1], Cyrille Bertelle[1], and Loay E. George[2]

[1]Normandie University, LITIS, FR CNRS 3638, ISCN, ULH, Le Havre, France

[2] Baghdad University, Computer Science Department, Baghdad, Iraq

*Abstract*

*The communication devices have produced digital traces for their users either voluntarily or not. This type of collective data can give powerful indications that are affecting the urban systems design and development. In this study mobile phone data during Armada event is investigated. Analyzing mobile phone traces gives conceptual views about individuals densities and their mobility patterns in the urban city. The geo-visualization and statistical techniques have been used for understanding human mobility collectively and individually. The undertaken substantial parameters are inter-event times, travel distances (displacements) and radius of gyration. They have been analyzed and simulated using computing platform by integrating various applications for huge database management, visualization, analysis, and simulation. Accordingly, the general population pattern law has been extracted. The study contribution outcomes have revealed both the individuals densities in static perspective and individuals mobility in dynamic perspective with multi levels of abstraction (macroscopic, mesoscopic, microscopic).*

## 1. INTRODUCTION

There's no doubt that mobile phone is the most dominant tool to gather human mobility data in spatio-temporal pattern due to its intense usage by human. Assuming every individual is represented by at least one mobile, even if he/she can use another communication method in view of great ICT invasion [1, 2, 3, 4, 5, 6]. The human mobility studies using CDRs has to be highlighted to find more modeling improvements, in order to have more realistic simulation results with flexible skills. In general, these studies are done basically in several concepts (Macroscopic, Mesoscopic, and Microscopic) of abstraction [7, 8, 9, 10].

The multi-agent simulations are very suggestive for spatio-temporal dynamics, since they are elaborating the relationships between micro-level individual actions and emergent macro-level phenomena as in [66]. According to [65], multi-agent system framework, which models emergent human social behaviors (competitive, queuing, and herding) at the microscopic level is used. This kind of models build artificial environment composed of agents, which have the ability to interact in intelligence and adaptability with each other [9, 12, 13]. In these models the agents are acting based on some strategies. However, their interactions are based on predetermined mobility conditions like (leader, follower, inhibition) agents. This type of simulation is very effective for large scale rescue scenario, complex systems, and modeling crowd behaviors [7, 14].

Recently, researches focusing on the study of the collective and individual human mobility using CDRs [44], the segmentation of urban spaces [67], and the understanding of social events [68] were done. The research [69] suggested the activity-aware map, using the user mobile phone to uncover the dynamic of inhabitants for urban planning, and transportation purposes. The researchers in [70] deal with large telecommunication database for Great Britain to explore human interactions, and emphasized on the highly interactions correlation with administrative regions. And in [71] they invented the co-location perspective to outline the behavior of the interacted telecom network users, who frequently call each other and share same spatio-temporal in a city. Whereas, the research [72] is presented the digital footprints (left traces), that are put through individuals during their mobility in urban space. This is performed using CDRs and georeferenced images of Flickr in order to investigate the tourists presence and mobility in Rome city. However, the research [73] is emphasized that urban mobility is highly correlated with individuals behavior during mobile phone usage at spatio-temporal patterns.

The human mobility understanding, modeling and simulating among urban regions is a challengeable effort, since it is very important for many kinds of events, either in the indoor events like evacuation of buildings, stadiums, theaters, ships, aircraft or outdoor ones like public assemblies, open concerts, religious gatherings, community evacuation ...etc. The simulation models of individuals mobility could be classified according to their space representation, which could be continuous; grid based or network structure. However, they could be classified according to their intent (specific, general), or the level of abstraction (macroscopic, mesoscopic, microscopic) [7, 8, 9, 10, 16, 17, 18]

The major limitation of CDRs data is the mass size, so they need the sampling and accurate manipulations to preserve their accuracy and to avoid any misused or misunderstanding. Another problem with the CDRs is the privacy problem, since the information of the mobile network subscribers will be vulnerable, so be announced for unauthorized parties; therefore it is anonymized (hidden) to protect the subscribers privacy. As well as, the graphical or geographical representation of these datasets is difficult, because of the hardware/software limitations that could be faced during processing. The CDRs datasets overcome the acquisition problems (financially, time consuming), but their projection and trajectories tracing still time and resource consuming. The researchers try to solve this limitation by selecting the samples randomly from the CDRs, and obtain the highly frequent records [60, 62, 30, 42, 63, 64, 32, 10, 33, 20, 12, 28].

This research captures the case study data in two phases; the analysis phase, and the modeling and simulation phase with regards to spatio-temporal features. The perspectives are elaborated in both static and dynamic views. The analysis and representation of the individuals concentration in the observed region are achieved in order to give a static view perspective for the investigated data, as will be shown in section 3. Thereafter, the statistical techniques are applied on the investigated data to model and simulate the human mobility, which gives the dynamic view perspective. The modeling and simulating of the Armada data event is performed at two levels:

1. Collective level: The data of aggregated space, aggregated time intervals, and both of them, which are pointing to aggregated data items. They are manipulated using statistical probability distributions. As will be elaborated in section 4.
2. Individual level: The data of space and time, which are pointing to individual items, they are used to reconstruct the individuals trajectories via a mathematical model of physical mobility characteristics. As will be elaborated in section 5.

## 2. DATA SET OF ARMADA CASE STUDY AND MOBILE NETWORKS DATA

The analysis of mobile phone traces allows describing human mobility with accuracy as never done before. These data are spatio-temporal data could be registered as CDRs (Call Detail Records), which have both time and spatial properties. Their unique features enable them to reflect human mobility in real time. Also, they have complex nature and a very rich environment to obtain human life pattern elaboration from multi points of view [19, 20, 4]. The main objective of this contribution is modeling and simulating individuals mobility during Armada event. The essential material used in this study is database generated by the given mobile phone operator (Orange Company in France) in form of CDRs.

Armada is the name of the famous marine festival takes ten days period, it takes place every 4 years periodically [21, 22]. Large sailing ships, private yachts, naval ships and even military frigate schooners are gathering from twenty different countries, in one of world free spectacle in Rouen (capital city of Upper Normandy in France), which is attracting 8 million visitors addition to Rouen citizens (estimated to 460,000 in 2008). This case study is observing the fifth edition of the Armada event during $4^{th}$ - $15^{th}$ July 2008. However, its miscellaneous activities are starting from 10:00 to 20:00 every day. The last day of this event coincides $14^{th}$ July (French national day) [21, 6]. The available data of Armada case study composed of 51,958,652 CDRs for 615,712 subscribers during 288 hours with lack of 15 hours, so the actual observed data of 273 hours. Armada event days are classified in two categories (days out of Armada $4^{th}$ and $15^{th}$ of July, and days within Armada $5^{th}$ - $14^{th}$ of July). During Armada period there are 5 days are off days, which are weekends and vacation, fall of (5, 6, 12, 13, and 14) of July, and other days (4, 7, 8, 9, 10, 11, and 15) of July which are work days. The Armada mobile phone database is composed of CDRs that contain: mobile IDs (alias), towers IDs and positions which are geo-referenced by 2D-coordinates (x, y), number of cells on each tower and Cells IDs, mobile activity types (call in/out, SMS in/out, mobile hand over, abnormal call halt, normal call end), and the date and time of the mobile phone activity recorded.

However, this kind of data is representing individuals occurrence in discrete (irrelevant) mode only, means that any mobile individual activity is recorded at (start/end) time, but there is lost information which is supposed to indicate the user's occurrence during inactive case (no data meanwhile the mobile phone is idle i.e. inactive or doesn't make any communication activities neither calls nor SMS activities). As well as, the available spatial data is only the towers (X, Y) coordinates. So, they would be considered to estimate individual transitions from position to another (from tower to tower), where each two consecutive activities between two towers could be considered as the displacement (mobility from coverage region to another), it is the main point to capture the transition from one location to another [25, 4]. These positions would be estimated approximately with regard to tower coverage (signal strength).

Seeing as these data have non-deterministic and discrete nature, therefore the collective mode would be the effective approach to be analyzed and simulated. As well as, since each individual could be disappeared for a while from the DB records which makes individual tracing is unworthy, and without significant indications on the individuals' mobility in the city. Nevertheless, this kind of analysis has two limitations, first one is the information given by mobile phone traces are incomplete to describe all individuals mobility. Actually, because many individuals moving in the city without using a mobile phone. However, the intensity of mobile phone usage nowadays helps to capture mobility analysis in an acceptable approximation. Second one is the mobile phone traces are only caught by a tower that is located at some specific places in the city, so the positions also would be estimated approximately. Consequentially, it is supposed to interpolate individual positions and reconstruct his trajectories from the traces to simulate his mobility (human mobility). The manipulation of these data requires several phases to get required analysis results and obtain precise expected knowledge about individuals densities and mobility [25, 26, 27, 2, 20, 12, 28].

Human mobility studies are benefiting from mobile phone data with concentration on the regions of higher mobile phone network coverage, these regions are characterized as (mature, stable, most developed). Under an assumption of that each mobile represents an individual, in other words his occurrence and mobility. As a consequence, individuals densities and mobility could be explored for the observed event.

## 3. DATA ANALYSIS AND VISUALIZATION

The GIS influences several scientific fields (urban planning, traffic management, sustainable infrastructure establishment, tourism… etc.). So, the combination of social sensor data with data mining techniques, and GIS could give deep total perception to urban dynamics towards socio-technical conceptualization for the city by elaborating the implicit daily patterns [29, 30, 10, 33, 34, 35, 5, 36, 37, 38, 11].

Specific platform has been developed to integrate and analyze the initial raw database; PHP server and SQL query language are used. Data visualization is accomplished using graphs, representing regions densities corresponding to human activities. The combination of these tools allows lining up the spatio-temporal data (CDRs) on the city map. ArcGIS platform is used for geographic representation, and to extract the aggregated data of five classified sectors, which are corresponding to the observed spatial sub-areas. The geographic region dividing considers sub-area of Armada activities as the central region and other sub-areas are its surroundings from four directions (north, south, east, and west). In order to summarize the spatial patterns, hence the study area is divided into adjacent sub-areas [38], this is done by using Voronoi polygons like in [39], where each polygon is associated with a tower and depicts the area under the main influence of this tower. Each Voronoi cell centering the tower coverage area, it corresponds all closest possible locations of individual detected in this coverage area. Each group of adjacent Voronoi polygons is formulating one of the five sub-areas. However, the center sector is located around Seine river banks where Armada event held, the eastern sector is located to the east of the center sector, the western sector is located to the west of the center sector, the northern sector is located to the north of center sector, and the southern sector is located to the south of the center sector. The overall perspective of the 30x30 Km observed region activities could be elaborated according to the daily intervals. As in figure 1, where the produced five sub-areas are presented [6].

The basic concept of this phase deals with analyzing and representation of individuals concentration in the observed region, according to (sub-areas during observed days, and sub-areas during observed hours). However, daily activity ratios of the 5 sectors are analyzed according to daily hours, and as obvious in figure 2.a. All sectors have similar patterns, but in varying ratios. The sectors could be ordered in descending order (center, south, north, west, and east). Whereas, highest activity ratios for all sectors are located in time intervals (00:00, 06:00-20:00), otherwise the activity ratios are descending over time intervals (01:00-06:00) and (21:00-23:00). The lowest activity ratios for all sectors are falling in time intervals (02:00-06:00), 15:00, and 23:00). All sectors have peak values in common 12:00 and 18:00 which are lunch hour, and end work hours respectively, where the social activities are in highest rank.

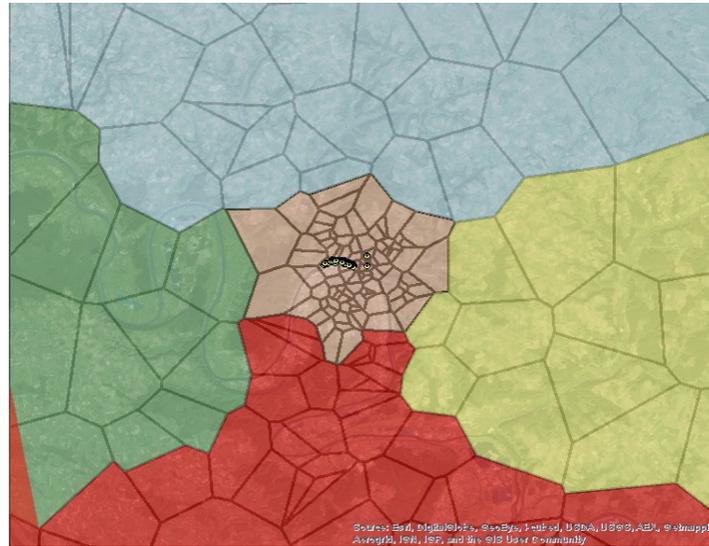

Figure 1: Densities analysis in macroscopic perspective: a. The five sub-areas appear in 5 colored zones, and are built by grouping Voronoi cells. The black anchors represent places along the Seine river platforms where held activities during Armada [6].

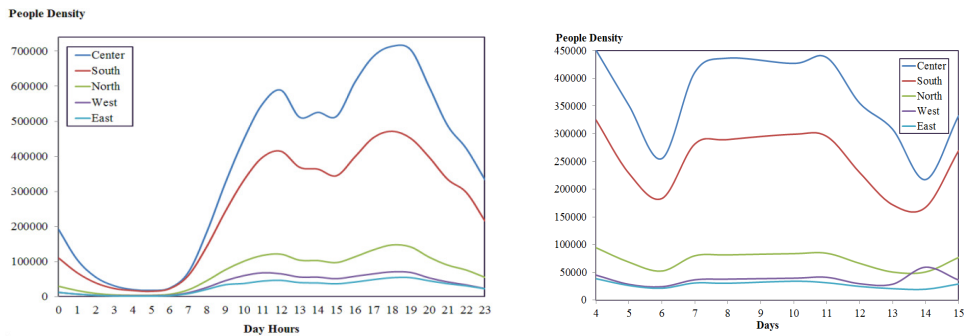

a.
b.

Fig. 2: Activity patterns according to days and hours: a: Daily analysis of the 5 sectors along 24 hours with average over the days period of Armada [6], b: Daily analysis of the 5 sectors along 12 days.

The most active sectors are the (center and south) along the observed period. There is high difference in activity ratios among sectors, so it could be ranked in descending order (center, south) as the most active sectors followed by (north, west, and south) sectors respectively. As well as, in figure 2.b the analysis of daily activity ratios for the 5 sectors along the 12 days. Where all sectors have highly activity ratios in the day $4^{th}$ July, since it is former to Armada event, hence the influence of all arrangements is clear. All sectors are decreasing in their activity ratios in the days (6, 12, and 13), which they are off days. That means the citizens are in their lowest activities during off days. All sectors have their high activity ratios in the day intervals (7-11) which they are work days. The anomalous event is appeared in day $14^{th}$ where all the sectors are in their lowest activity ratios except the west sector where its activity ratios mark peak values. This could be interpreted as France National Day influences this sector, the ceremonial may be held in this sector during this day [6]. This phase considers the phones activities as aggregated individuals (densities) in aggregated geographic locations. Where each mobile phone could represent an individual, and each group of adjacent Voronoi polygons represents one geographic sub-area, that are formulated the total observed region.

## 4. DESCRIBING INDIVIDUALS' ACTIVITIES

The simulation process reflects the characteristics and behavior of any system. The word (simulation) is used in several fields it includes the modeling of natural sentences or human organs as an attempt to explore the details of this process. However, in this phase there is an attempt to model and simulate the general pattern law for the collective data, and then they are investigated in more detail by classifying them into work and off days. The mobile phone data are very heterogeneous nature, since the users could be very active having many calls/SMSs, or inactive having little usage of the network, so sampling the users would be depending on their activities number [55]. Since, the individuals are varied in their usage of mobile phone, which they are ranged between (rarely-frequently) usages during observed period. Therefore, the individuals are grouped according to their total activities. However, the probability distribution of waiting time (inter-event time $\Delta T$) has been computed for each consecutive activity and for each individual. In order to formulate simplified models with their quantitative analysis parameters, thus the probability function should be computed to obtain the general system pattern, which is done according to consecutive inter-event time parameter. The probability distribution of individuals activities are made by [40, 41, 42, 43, 44, 50]:

1. Compute probability function to get the system universal behavior (pattern), according to consecutive inter-event time.
2. Compute the probability density function to get each agent sample probability [48, 43].

The individuals are grouped with regard to their activities, as in figure 3.a, which it explores the long waiting times are characterizes the individuals of less activities. The waiting time distribution is also computed for the work days as in figure 3.b, and then it is computed for the off days as in figure3.c. However, the distribution of the inert-event time $\Delta T$ is estimated by equation (2):

$$P(\Delta T) = (\Delta T)exp^{-(\Delta T)} \ldots \ldots (2)$$

### 4.1. Fixed Inter-Event Time Observations

The modeling and simulation process starts by computing probability model, which is modeled the data and simulated the responses of the model by using one of the well-known functions probability density function PDF [50, 51].

The main parameters in this study are $\Delta t$, mean of $\Delta t$, and $r_g$. The samples means of the distributions are drawn from an exponential distribution. Then, compute the mean of all means. However, understand the behavior of sample means from the exponential distribution in order to have the universal system pattern (general population law). The computations are done sequentially by manipulating all 12 days data (each day independently) for total individuals in spatio-temporal manner, and then eliminate the individuals of the only one occurrence in the CDRs, since they didn't have a significant indication on mobility. As well as, sorting the data by time in order to have the real sequence of positions transitions of individuals trajectories, and classify the data according to individuals' activities (sampling), by computing $\Delta t$ (inter-event time/ waiting time), where it is the time elapsed between each successive activities of each individual. It ranges between 15-1440 minutes, this sampling is done according to logical intuition, since 15 minutes is the minimum time that can give mobility indication, and the 1440 (24 hours) could be considered as highest elapsed time to travel in the city, then compute $\Delta T_a$ (average inter-event time) of all individuals, so classify (min, max) samples according to activities score (activities densities). Thereafter, sort the samples of activities according to inter-event time, and then compute the inter-event time of all individuals $\Delta t$ and $\Delta T_a$ for each sample,

36

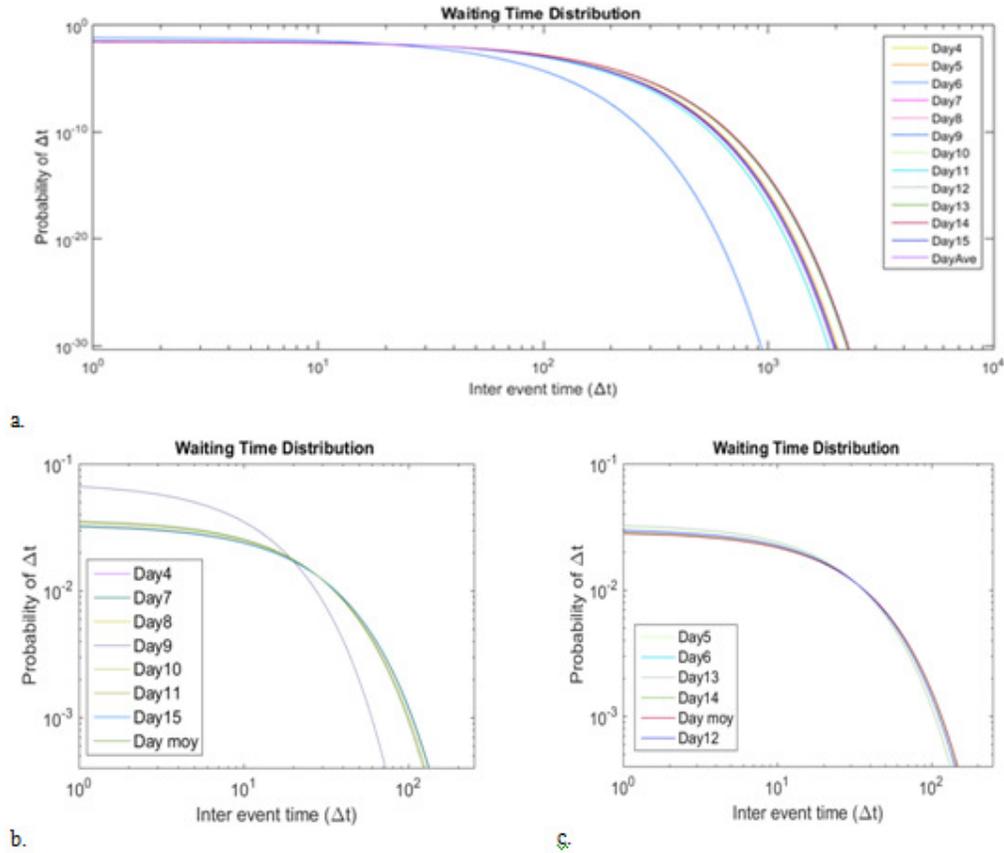

Fig. 3: Waiting time distribution in mesoscopic perspective: a: Waiting time distribution $p(\Delta T)$ of mobile phone activities, where $\Delta T$ is the spent time between each two successive activities, legend symbols are used to distinguish the unique days and the curve of their mean (DayAve)for the total population during the total period [78], b: Work days waiting time distribution p($\Delta$t) of mobile phone activities, legend symbols are used to distinguish the unique days and the curve of their mean (Daymoy) for the total population during the total period, c: Off days waiting time distribution p($\Delta$t) of mobile phone activities, legend symbols are used to distinguish the unique days and the curve of their mean (Daymoy) for the total population during the total period.

where ($\Delta$t / $\Delta T_a$) is the average inter-event time of the total individuals (population) as $\Delta T_a = 1431\ minutes$, then compute exponential distribution probability for the total data. Accordingly, identify the general pattern law of the population.

Exponential distribution (probability distribution) is capable of modeling the events happened randomly over time. In this case, it is able to describe the inter-event time and the average of inter-event time [47, 54, 51, 53] of individuals' activities with PDF (Probability Density Function) as in equation (3). However, the cutoff distribution is determined by the maximum observed inter-event time at which the individual can wait to make any mobile activities, where it is $\Delta t = 1431\ minutes$. The developed algorithm to compute probability distributions of activities and displacements by inter-Event Times has $O(n^3 + 2n^2 + n)$ complexity, and it is implemented with Matlab2015 platform.

$$f(x|\mu) = \frac{1}{\mu} e^{\frac{-x}{\mu}} \quad \ldots\ldots (3)$$

As well as, the displacement statistics $p(\Delta r)$ has been computed for the total individuals (population) during the observed period. As in figure 4.a the displacements distribution is computed. Also, it is computed for work days as in figure 4.b, and computed for off days as in figure 4.c. However, the Δr is the covered distance between each two successive activities during time Δt, where it is in the range $20 - 1440\ minutes$, the investigated distance would be limited by the maximum distance that could be traveled by individuals in the $\Delta T$. Hence, the cutoff distribution is determined by the maximum observed distance at which the individual can travel, where it is $\Delta r = 7.23e + 04$ m along day hours, since the maximum time slice couldn't exceed $24\ hours$ with regards to observed region. The consequence is the $p(\Delta r)$ distributions for different $\Delta t$ follows truncated power law, as in equation (4).

$$P(\Delta r) = (\Delta r)exp^{-(\Delta r)} \quad \ldots\ldots (4)$$

The distribution of $r_g$ uncovers the population heterogeneity, where individuals traveled in $P(r_g)$ in (long/short) distances regularly within $r_g(t)$ as formulated in equation (5). which refers to the center of mass of the individual trajectory.

$$r_g^a = \sqrt{\frac{1}{n_c^a(t)} \sum_{i=1}^{n_c^a} (\vec{r_i^a} - \vec{r_{cm}^a})^2} \quad \ldots\ldots (5)$$

Where $\vec{r_i^a}$ refers to i=1... $n_c^a(t)$ positions recorded for individual a, and $\vec{r_{cm}^a} = \frac{1}{n_c^a(t)} \sum_{i=1}^{n_c^a} \vec{r_i^a}$,

The distribution $P(r_g)$ produces power law investigated in the aggregated traveled distance (displacements) distribution $P(r_g)$ as in equation (6).

$$p(r_g) = (r_g)exp^{-(r_g)} \quad \ldots\ldots (6)$$

The developed algorithm to compute radius of gyration distribution has $O(n^2 + 4n)$ complexity, and it is implemented with Matlab2015 platform. Whereas, the $p(r_g)$ is computed for total population along observed period, in order to classify the $r_g$ evolution along the time series, as figure 5.a shows the $r_{g_s}$ variance along time, they are approximately similar in their patterns, also the highest $r_{g_s}$ are the small once ranged $(5 - 20)$, while the range $(25 - 30)$ are the lowest once to indicate that individuals who have the tendency to mobile in small $r_{g_s}$ and in stable patterns along observed period. As well as, the distribution of $r_{g_s}$ along time is computed for the data of work days and for off days as in figures 5.b, 5.c respectively.

### 4.2. Trajectories in Intrinsic Reference Frame

The significant importance of revealing human trajectories enforces the tendency to build the statistical models. Human trajectories have random statistical patterns; hence tracing human daily activities is the most challengeable issue, in addition to its importance, which is mentioned earlier as urban planning, spread epidemics... etc. In spite of data sources variance (billing system, GSM, GPS), but the common characteristics are the aggregated jump size (Δr), and waiting time (Δt) distributions. Where, (Δr) gives an indication on the covered distances by an individual in (Δt) for each two consecutive activities, and the (Δt) is the time spent by the individual between each two consecutive activities [53]. The individual's trajectory considered as the microscopic level of mobility abstraction, which is constituted of sequenced coordinates positions along the time, i.e. the agent displacement in spatio-temporal unit. Therefore, the consecutive phone activities are good proxies to compute individuals travel distances (displacements) [30, 53, 54, 57, 35]. These distributions are overlapped for groups, however the radius of gyration ($r_g$) is considered to be a more dedicated feature that capable of characterizing the travel distances ($\Delta$r) of individuals. The distributions showed that the ($\Delta$r) of individuals are almost identical, where the periodic trajectories are invariant. As well as, the activities distributions are uncovered the similarities of regular patterns, during the time evolution of radius of gyration ($r_g$). The experiments revealed the possibility of classifying individuals activities into some patterns samples. As well as, almost individuals have similar activities patterns, and these activities in general could be varied depending on the days either working or off days.

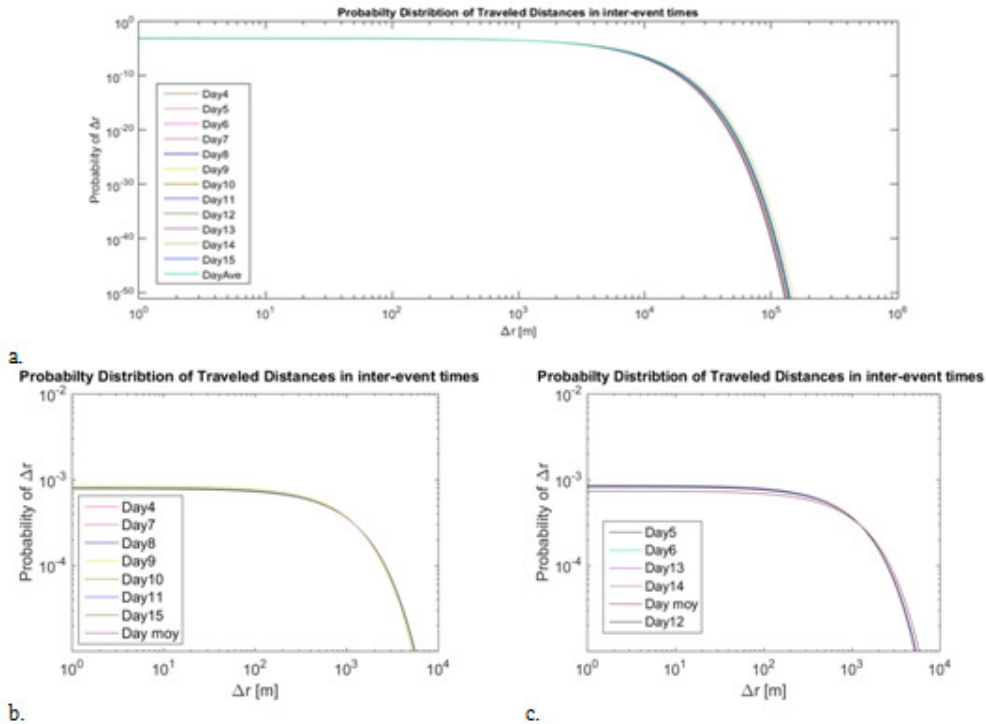

Fig. 4: Displacements distribution in mesoscopic perspective: a: Displacements distribution $p(\Delta r)$ for waiting times ($\Delta t$), cutoff distribution is determined by the maximum distance traveled by individuals for specific inter-event-times, each day has its own curve and the curve of days mean is (DayAve) [77], b: Work days displacements distribution p($\Delta$r) for waiting times ($\Delta$t), cutoff distribution is determined by the maximum distance traveled by individuals for specific inter-event-times, each day has its own curve and the curve of days mean is (Daymoy), c: Off days displacements distribution p($\Delta$r) for waiting times ($\Delta$t), cutoff distribution is determined by the maximum distance traveled by individuals for specific inter-event-times, each day has its own curve and the curve of days mean is (Daymoy).

## 5. INDIVIDUAL TRAJECTORY CHARACTERISTICS

The individuals activities sparseness causes incomplete spatial information, therefore the mobile individual has some general physical characteristics, which are useful to compensate the lack of data (when no activity recorded), and that could be used to build a mathematical model of human mobility patterns. In order to reconstruct the individual trajectory even if there is no mobile activity (no data), so the most common mobility characteristics are used. They are as follows [55, 56, 57, 58, 32, 61, 24, 59]:

1. Center of mass: It's the most visited positions by individual $c_m$, as in equations (7-8) respectively:

$$x_{cm} = \sum_{i=1}^{n} x_i/n \quad \ldots\ldots (7)$$

$$y_{cm} = \sum_{i=1}^{n} y_i/n \quad \ldots\ldots (8)$$

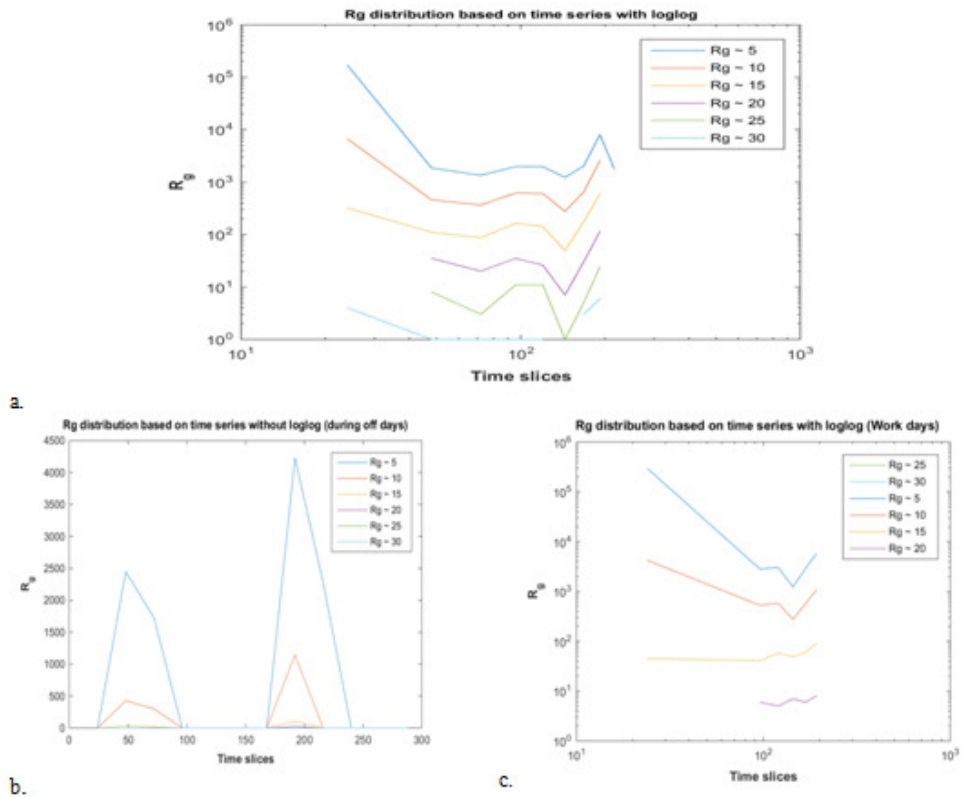

Fig. 5: Radius of gyration distribution in mesoscopic perspective: a: $R_g$ Distribution based on time series during observed period, b: Rg distribution based on time series during work days period, c: Rg distribution based on time series with loglog during off days period.

Where $x_i$ and $y_i$ are the coordinates of the spatial positions, $n$ is the number of spatial positions, that are recorded in the CDRs.

2. Radius of Gyration: It is the average of all individual's positions, which is the indication of the area visited by the individual (traveled distance during the time ) as in equation (5)
3. Most frequent positions to uncover the individual tendency, according to visited locations.
4. Principal axes $\theta$ (moment of inertia): This technique makes it possible to study some individuals trajectories in a common reference frame, by diagonalizing each of trajectory

40

Informatics Engineering, an International Journal (IEIJ), Vol.4, No.2, June 2016inertia tensor, hence to compare their different trajectories. Moment of inertia to any object is obtained from the average spread of the mass of an object from a given axis. This could be elaborated using two dimensional matrix called tensor of inertia. Then, by using standard physical formula, the inertia tensor of individual's trajectories could be obtained, as in the equations (9-14) respectively:

$$I = \begin{pmatrix} I_{xx} & Ixy \\ Iyx & Iyy \end{pmatrix} \quad \ldots \ldots (9)$$

$$I_{xx} = \sum_{i=1}^{n} y_i^2 \quad \ldots \ldots (10)$$

$$I_{yy} = \sum_{i=1}^{n} x_i^2 \quad \ldots \ldots (11)$$

$$I_{xy} = I_{yx} = -\sum_{i=1}^{n} x_i y_i \quad \ldots \ldots (12)$$

$$\mu = \sqrt{4 I_{xy} I_{yx} + I_{xx}^2 - 2 I_{xx} I_{yy} + I_{yy}^2} \quad \ldots \ldots (13)$$

Obtaining the following equation (19):

$$\cos \theta = -I_{xy} (1/2 I_{xx} - 1/2 I_{yy} + 1/2 \mu)^{-1} \frac{1}{\sqrt{1 + \frac{I_{xy}^2}{(1/2 I_{xx} - 1/2 I_{yy} + 1/2 \mu)^2}}} \quad \ldots \ldots (14)$$

Rotation by θ produced a conditional rotation of 180° as the most frequent position lays in $x > 0$.

5. **Standard Deviation**: To verify the horizontal and vertical coordinates of individual mobility in the intrinsic reference frame. However, trajectories are scaled on intrinsic axes using standard deviation of the locations for each individual a, as in equations (15-16) respectively.

$$\sigma_x^a = \sqrt{\frac{1}{n_c^a} \sum_{i=1}^{n_c^a} (x_i^a - x_{cm}^a)^2} \quad \ldots \ldots (15)$$

$$\sigma_y^a = \sqrt{\frac{1}{n_c^a} \sum_{i=1}^{n_c^a} (y_i^a - y_{cm}^a)^2} \quad \ldots \ldots (16)$$

Then, obtain the universal density function using the following equation (18):

$$\tilde{\phi} = (x/\sigma_x, y/\sigma_y) \quad \ldots \ldots (18)$$

By using the spatial density function to aggregate the individuals with the common $r_{g_s}$. Then, individuals had been chosen from different classified groups with regard to their radius of gyration, the trajectories would be rescaled for a group of individuals according to the mobility characteristics mentioned above. The potential trajectory of the individual has been made to visualize and analyze the mobility of 3 individuals during observed period. However, the individuals had been chosen, according to their pertinence of $r_g$, where each one is elected randomly from the unique $r_g$ sample, then the potential trajectories of the individuals ($individual_1$, individual, $individual_3$) are chosen from $R_{g_2}$, $R_{g_9}$, $R_{g_{15}}$ respectively, then computed and simulated in off days as in figure 6 and in work days as in and figure 7 respectively.

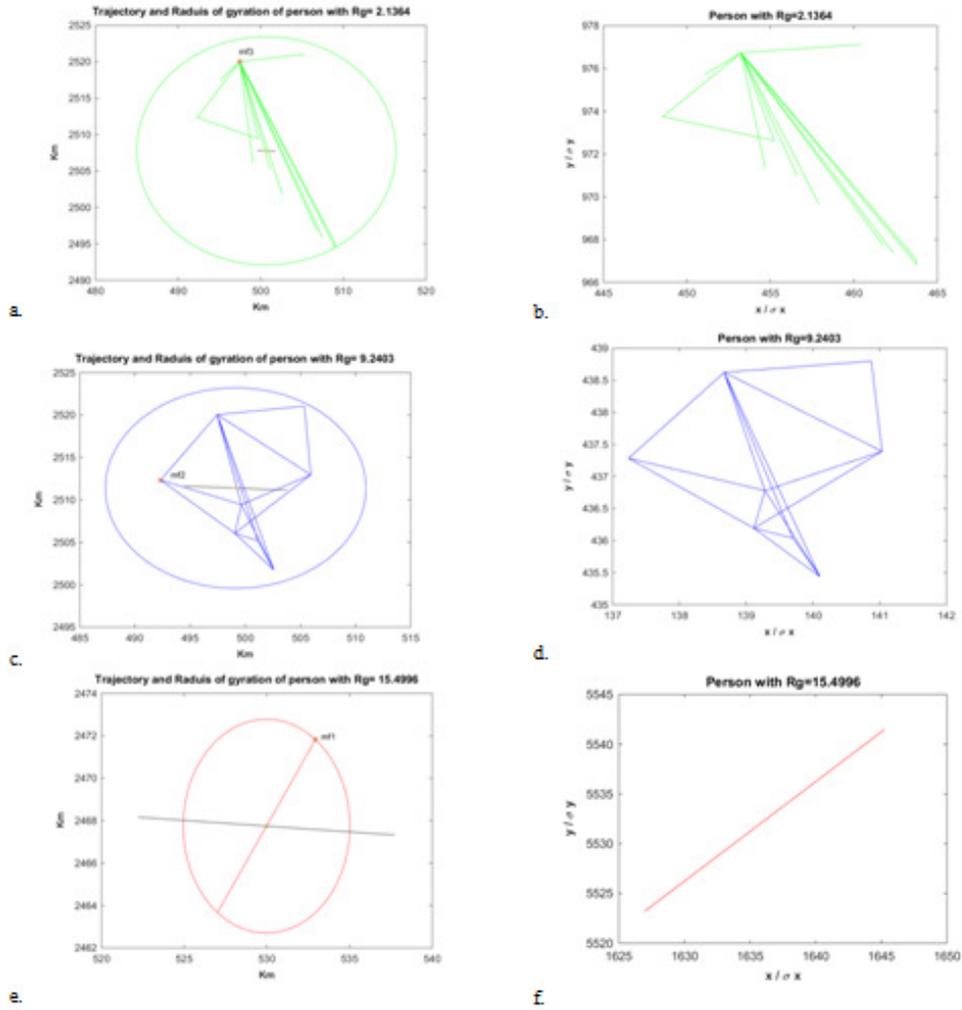

Fig. 8: Estimated individuals' trajectories within intrinsic reference frame in microscopic perspective during off days with red circle refers to most frequent position: a: Individual trajectory chosen from $R_{g_2}$, b: individual scaled trajectory chosen from $R_{g_2}$, c: Individual trajectory chosen from $R_{g_9}$, d: Individual scaled trajectory chosen from $R_{g_9}$, e: Individual trajectory chosen from $R_{g_{15}}$, f: Individual scaled trajectory chosen from $R_{g_{15}}$.

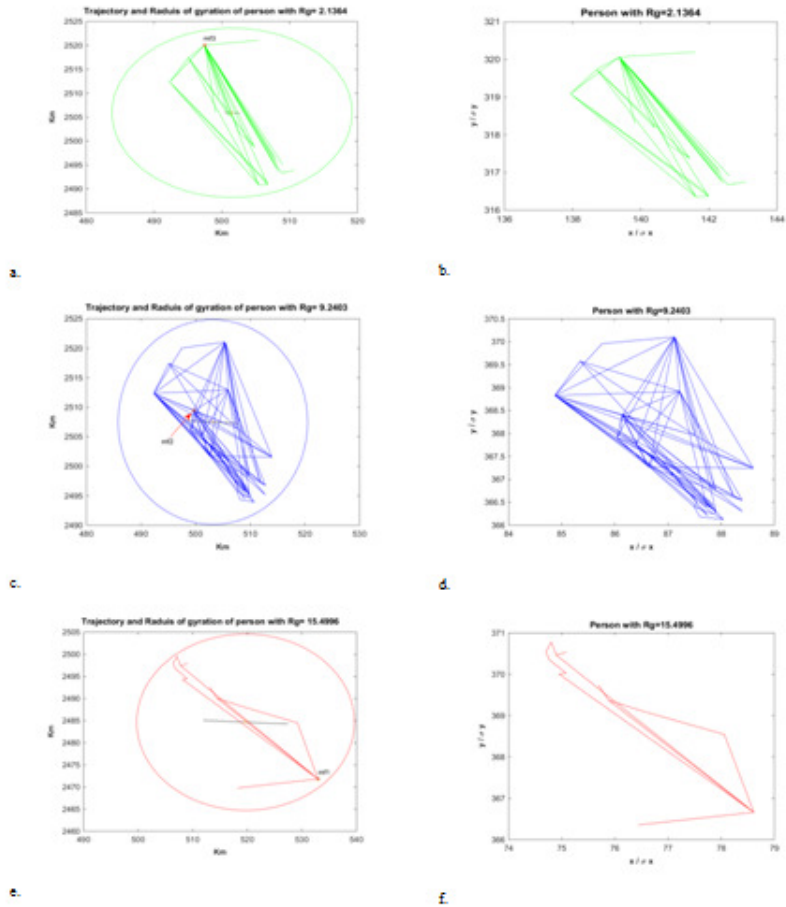

Fig. 9: Estimated individuals' trajectories within intrinsic reference frame in microscopic perspective during work days with red circle refers to most frequent position: a: Individual trajectory chosen from $R_{g_2}$, b: individual scaled trajectory chosen from $R_{g_2}$, c: Individual trajectory chosen from $R_{g_9}$, d: Individual scaled trajectory chosen from $R_{g_9}$, e: Individual trajectory chosen from $R_{g_{15}}$, f: Individual scaled trajectory chosen from $R_{g_{15}}$.

## 6. Conclusions

This research aimed to understand human mobility using the CDRs. The investigation endorsed that mobile phone activities can reflects individuals density, also could act as a feature of grouping the total number of mobile network users, hence each group will have its own feature. Individuals density and dense variance between the sub-areas could be estimated in order to explore the tendency or the orientation of individuals in these areas, but almost in a static manner.

As well as, it is concluded that the most common parameters of modeling human mobility (inter-event time ∆t, travel distance (displacement) ∆r and radius of gyration $r_g$) are represented by the power-law distribution. These parameters can reveal mobility patterns of the evolved population.

This study confirmed that radius of gyration ($r_g$) is the most common quantity, which is associated with human mobility trajectories, due to its capability in measuring the how far the mass from the center of mass. However, in this case study it is gradually increased at the beginning, but it settles down versus time. It has a key effect on the travel distance (Δr) distributions.

It is recommended that further research be undertaken in the following areas: Perform more mobility analysis for extracting new human mobility parameters to make a more detailed description for individuals behavior. Use more comprehensive statistical methods to construct human models for the society via CDRs. Analyze more CDRs data for the same region, but in time periods other than Armada period, so the comparison between the two cases can be more useful to describe the Armada impact on the city.

Informatics Engineering, an International Journal (IEIJ), Vol.4, No.2, June 2016